\documentclass[
aps,
twocolumn,
superscriptaddress,
prl,
showkeys,
showpacs,
nofootinbib,
notitlepage,
floatfix,
]{revtex4-2}

\usepackage{xcolor}
\usepackage{amssymb}
\usepackage{graphicx}
\usepackage{subfigure}
\usepackage{bm}
\usepackage[T1]{fontenc}
\usepackage{lineno}
\usepackage{xspace}
\usepackage{booktabs}
\usepackage{tabularx}
\usepackage{array}
\usepackage{siunitx}
\usepackage{dcolumn}
\usepackage{amsmath}
\usepackage{wasysym}
\usepackage[colorlinks=true,citecolor=blue,filecolor=blue,linkcolor=blue,urlcolor=blue,pdftex]{hyperref}
\usepackage{orcidlink}
\usepackage{currfile}

\begin{document}
%\linenumbers
\title{Searching for Dark Photons with a room-temperature dielectric haloscope}

\newcommand{\westlake}{\affiliation{Department of Physics, School of Science, Westlake University, Hangzhou 310030, China}}

\author{Siyin~Li\,\orcidlink{0009-0008-7204-8093}}\affiliation{Department of Physics, School of Science, Westlake University, Hangzhou 310030, China}
\author{Paschos~Ioannis\,\orcidlink{0000-0001-9762-6971}}\affiliation{Department of Physics, School of Science, Westlake University, Hangzhou 310030, China}
\author{Zhengyin~Yang\,\orcidlink{0009-0003-2047-350X}}\affiliation{Department of Physics, School of Science, Westlake University, Hangzhou 310030, China}
\author{Alexandros Spiliotis\,\orcidlink{0000-0003-4051-4439}}
\affiliation{Department of Physics, School of Science, Westlake University, Hangzhou 310030, China}
\author{Husheng~Guan\,\orcidlink{0009-0006-5049-0812}}\email[]{guanhusheng@westlake.edu.cn}\affiliation{Department of Physics, School of Science, Westlake University, Hangzhou 310030, China}
\affiliation{Institute of Natural Sciences, Westlake Institute for Advanced Study, Hangzhou 310024, China}
\author{Pavlos~G. Savvidis\,\orcidlink{0000-0002-8186-6679}}\email[]{p.savvidis@westlake.edu.cn}
\affiliation{Department of Physics, School of Science, Westlake University, Hangzhou 310030, China}
\author{Shengchao~Li\,\orcidlink{0000-0003-0379-1111}}\email[]{lishengchao@westlake.edu.cn}
\affiliation{Department of Physics, School of Science, Westlake University, Hangzhou 310030, China}

\date{\today}

\begin{abstract}
We present a search for dark-photon dark matter with a room-temperature dielectric multilayer haloscope. The dielectric stack enhances photon conversion near 2\,eV, and a spatially resolved CMOS focal plane records the emitted photons with few-photon sensitivity. We calibrate the stack--lens--CMOS response \textit{in situ} and use the calibrated focal-plane pattern in a template-based inference. With 904\,h of search data and 404\,h of background-control data, we observe no excess and set a 90\% confidence-level upper limit of $\kappa < 4.0\times10^{-13}$ for dark-photon dark matter with mass 1.9$\,\mathrm{eV}/c^2$.
\end{abstract}


\maketitle

\textit{Introduction.---}Dark matter (DM) remains a central open problem in particle physics and cosmology. Ultralight bosonic fields, including axions, axion-like particles~\cite{Arias:2012az,Graham:2015ouw}, and dark photons~\cite{Fabbrichesi:2020wbt,Caputo:2021eaa}, offer well-motivated candidates with distinctive laboratory signatures. The dark photon, a gauge boson associated with a hidden $U_{\chi}(1)$ symmetry~\cite{Holdom:1985ag}, can kinetically mix with the Standard Model (SM) photon and thereby couple to electromagnetic detectors. Dark-photon dark matter (DPDM) can arise from several early-universe mechanisms, including enhanced misalignment production in the presence of kinetic mixing or mass thresholds~\cite{Nelson:2011sf,Redondo:2008ec,Dubovsky:2003yn,Abel:2008ai}, quantum fluctuations during inflation~\cite{Ema:2019yrd,Dimopoulos:2008rf,Caldwell:2011ra,Ford:1986sy}, parametric resonance after inflation~\cite{Co:2018lka,Bastero-Gil:2018uel,Kofman:1997yn,Greene:1997fu,Agrawal:2018vin}, and topological-defect decay, particularly from cosmic strings~\cite{Long:2019lwl,Vilenkin:1982hm}.

If dark photons constitute all or part of the DM, they act as a coherent oscillating field with frequency set by their mass. This motivates laboratory searches for DPDM through its coupling to the electromagnetic current~\cite{Horns:2012jf}. At low energies, a terrestrial detector immersed in this nonrelativistic background is described by the effective Lagrangian

\begin{align}
\mathcal{L} \supset & -\frac{1}{4}F_{\mu\nu}F^{\mu\nu} - \frac{1}{4}F'_{\mu\nu}F'^{\mu\nu} + \frac{\kappa}{2}F_{\mu\nu}F'^{\mu\nu} \notag \\
& + \frac{m^2_{A'}}{2}A'_{\mu}A'^{\mu} + e J^{\mu} A_{\mu},
\end{align}
where $F_{\mu\nu}$ and $F'_{\mu\nu}$ are the field-strength tensors of the photon and dark photon, $\kappa$ is the kinetic-mixing parameter, and $A'_{\mu}$ is the dark-photon field with mass $m_{A'}$. The last term describes the interaction of photons with the four-current $J^{\mu}$.

A variety of resonant and broadband techniques target this signal, including microwave cavities~\cite{SHANHE:2023kxz,APEX:2024jxw}, LC circuits~\cite{Ghosh:2021ard}, and quantum sensors such as transmon qubits~\cite{Kang:2025kaf}, atomic magnetometers~\cite{Xu:2023vfn}, broadband cavity arrays~\cite{Chen:2023ryb}, and quantum upconversion schemes~\cite{Agrawal:2023xxb}. Sensitivity typically degrades toward optical frequencies, corresponding to eV-scale masses, where resonant enhancement and ultralow-noise readout become challenging~\cite{Chen:2023ryb,Liu:2024puo}. If inflationary fluctuations generate the observed DPDM relic abundance, a detection at eV-scale masses would point to an inflationary Hubble scale of order $10^{12}$\,GeV~\cite{Graham:2015rva,Bastero-Gil:2018uel}, a scale not directly accessible to collider experiments.

Recent studies identify viable direct-detection targets for DPDM over a broad mass range, including the eV scale~\cite{Cyncynates2025Targets,Cyncynates2025DefectFree}. We target this range with an alternating thin-film dielectric stack operated at room temperature. In a periodic dielectric multilayer, the reciprocal lattice vector compensates the momentum mismatch between the nonrelativistic DPDM field and an on-shell photon, so radiation from successive interfaces adds coherently~\cite{Baryakhtar2018}. A stack satisfying the half-wave resonance condition enhances dark-photon-to-photon conversion at the target frequency~\cite{Baryakhtar2018}. For virialized halo DPDM, the emitted photons inherit the dark-photon energy and emerge within a narrow cone of opening angle $\theta\sim v_{\rm DM}/c\sim10^{-3}\,\mathrm{rad}$ around the surface normal~\cite{Baryakhtar2018,Manenti:2021whp}. This near-normal emission can be collected with compact optics and imaged on a low-noise, single-photon-sensitive focal plane.

The optical-haloscope geometry also predicts a localized focal-plane light distribution rather than only a total photon rate~\cite{Baryakhtar2018,Manenti:2021whp}. A calibrated spatial template therefore provides a stricter consistency check for any candidate excess, while the wavelength-dependent stack response offers a route to extract dark-photon mass information in future multi-stack measurements. CMOS sensors are well suited to this implementation because they provide large-format imaging, mature on-chip functionality, and intrinsically multiplexed readout~\cite{Bigas2006}. This use of resolved detector information is consistent with the broader strategy of low-background pixel searches, including skipper-CCD searches such as SENSEI~\cite{PhysRevLett.125.171802}.

Our approach, Spatially-resolved Photon Emission Conversion with Transmission Reflector Arrays (SPECTRA), implements a dielectric optical haloscope with CMOS imaging readout. The present prototype integrates a 47-pair dielectric stack, a focusing lens, and a cooled CMOS sensor into a compact room-temperature detection platform.
A DPDM signal is tested through both its total photon yield and its calibrated focal-plane light pattern, determined by the stack response, optical geometry, and detector alignment. We obtain this template from \textit{in situ} laser calibration and optical simulation, and incorporate it into the statistical inference, yielding an imaging-based search for eV-scale DPDM.

\begin{figure}[!tbp]
    \includegraphics[width=\linewidth]{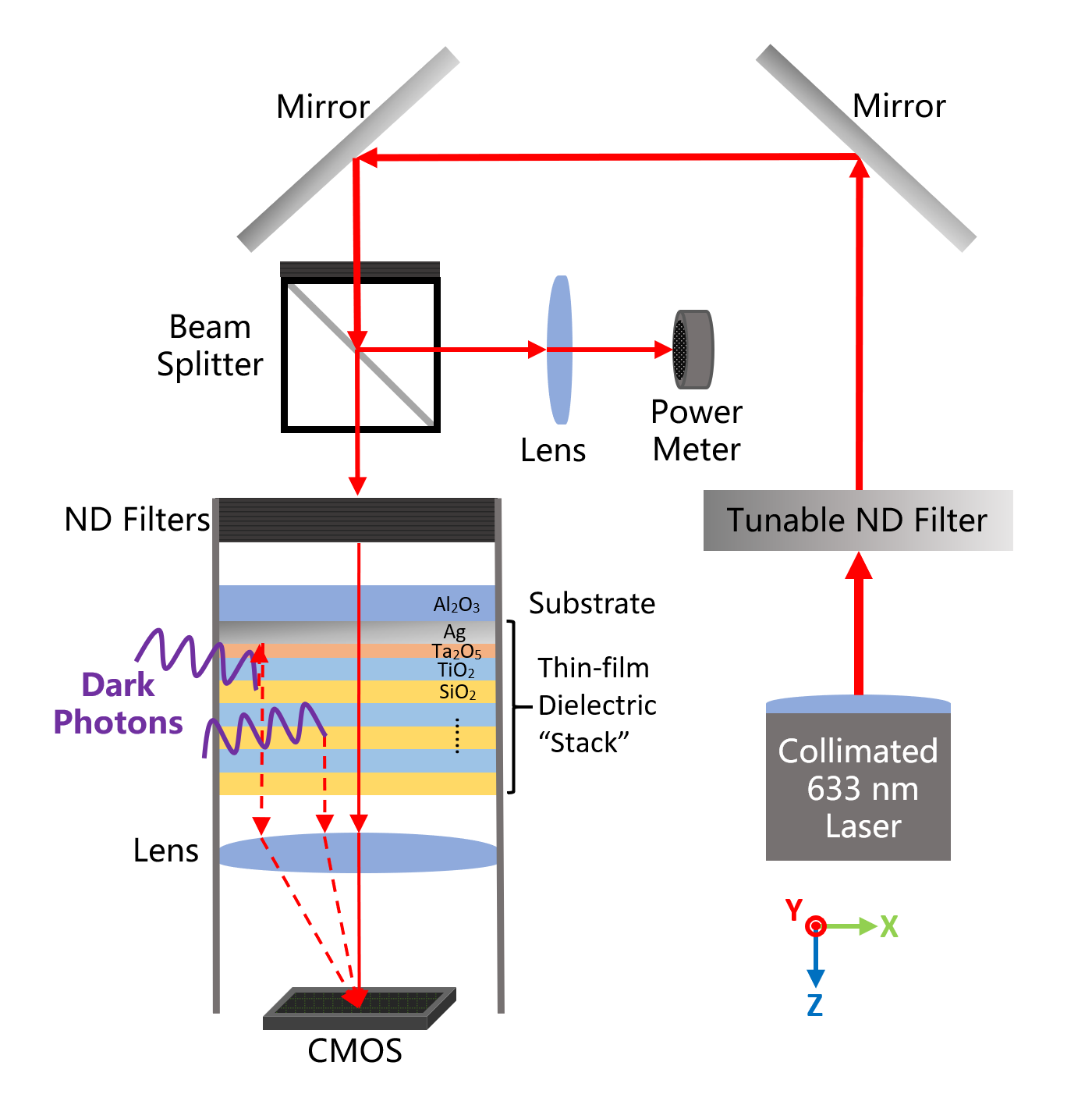}
    \caption{Schematic of the SPECTRA prototype. The core detector (lower left) consists of a 1-inch cylindrical dielectric stack with 47 alternating TiO$_2$/SiO$_2$ pairs, a focusing lens, and a CMOS sensor. We house the assembly inside a lens tube and cover the optical opening with neutral-density filters as needed for calibration. The external calibration system, shown not to scale, includes a collimated 633~nm He--Ne laser, a variable attenuator, a beam-splitter power monitor with a mounted filter, and two steering mirrors. The \textit{x}, \textit{y}, and \textit{z} axes are indicated.}
    \label{fig:setup}
\end{figure}

\textit{Experimental setup.---}The SPECTRA prototype contains two optical chains. The \textit{primary} chain consists of a multilayer dielectric stack, a best-form focusing lens, and a cooled CMOS imaging sensor, as sketched in Fig.~\ref{fig:setup}. We mount these components inside a 1-inch black-anodized aluminum lens tube to maintain light tightness and mechanical stability. The assembled system is placed on an optical table inside a light-tight enclosure, which provides additional shielding and supports laser calibration.
We fabricate the stack on a 1-inch epi-polished sapphire substrate with thickness 430\,$\mu$m; the Supplemental Material~\cite{suppmat} gives the fabrication and characterization details. The stack consists of alternating TiO$_2$ and SiO$_2$ layers, each designed to satisfy the half-wave optical-thickness condition, $nd\approx317$ nm. This design maximizes the conversion near 633 nm and matches the wavelength of the calibration laser. A 1-inch, 75-mm-focal-length lens focuses the near-normal emitted photons onto the downstream sensor.

The photosensor is a 16-mm-format Sony IMX533CLK-D CMOS device comprising a $3000 \times 3000$ pixel array with a pixel pitch of $3.76~\mu\mathrm{m}$. Thermoelectric cooling maintains the sensor at $-25^\circ\mathrm{C}$, reducing dark current and readout noise sufficiently to enable single-photon-sensitive operation. Although the manufacturer-provided quantum-efficiency curve indicates a value near $77\%$ at $633~\mathrm{nm}$ ($474~\mathrm{THz}$, denoted $\omega_0$), we determine the absolute stack--lens--CMOS detection efficiency at $\omega_0$ from \textit{in situ} calibration and use the datasheet curve only to model the relative wavelength dependence.

The \textit{secondary} optical chain is used for \textit{in situ} calibration and alignment. A fiber collimator delivers a stable 633-nm beam to the stack. A variable neutral-density (ND) filter sets the intensity for focusing and weak-signal tests, while two broadband mirrors adjust the incidence angle and beam position. For hour-long exposures in the single-photon regime, fixed 1-inch ND filters are installed inside the lens tube and beam-splitter cage, providing 16 orders of magnitude of attenuation before the beam reaches the stack. Laser stability during extended calibration runs is monitored with a downstream power meter.

\begin{figure}[!tbp]
    \includegraphics[width=\linewidth]{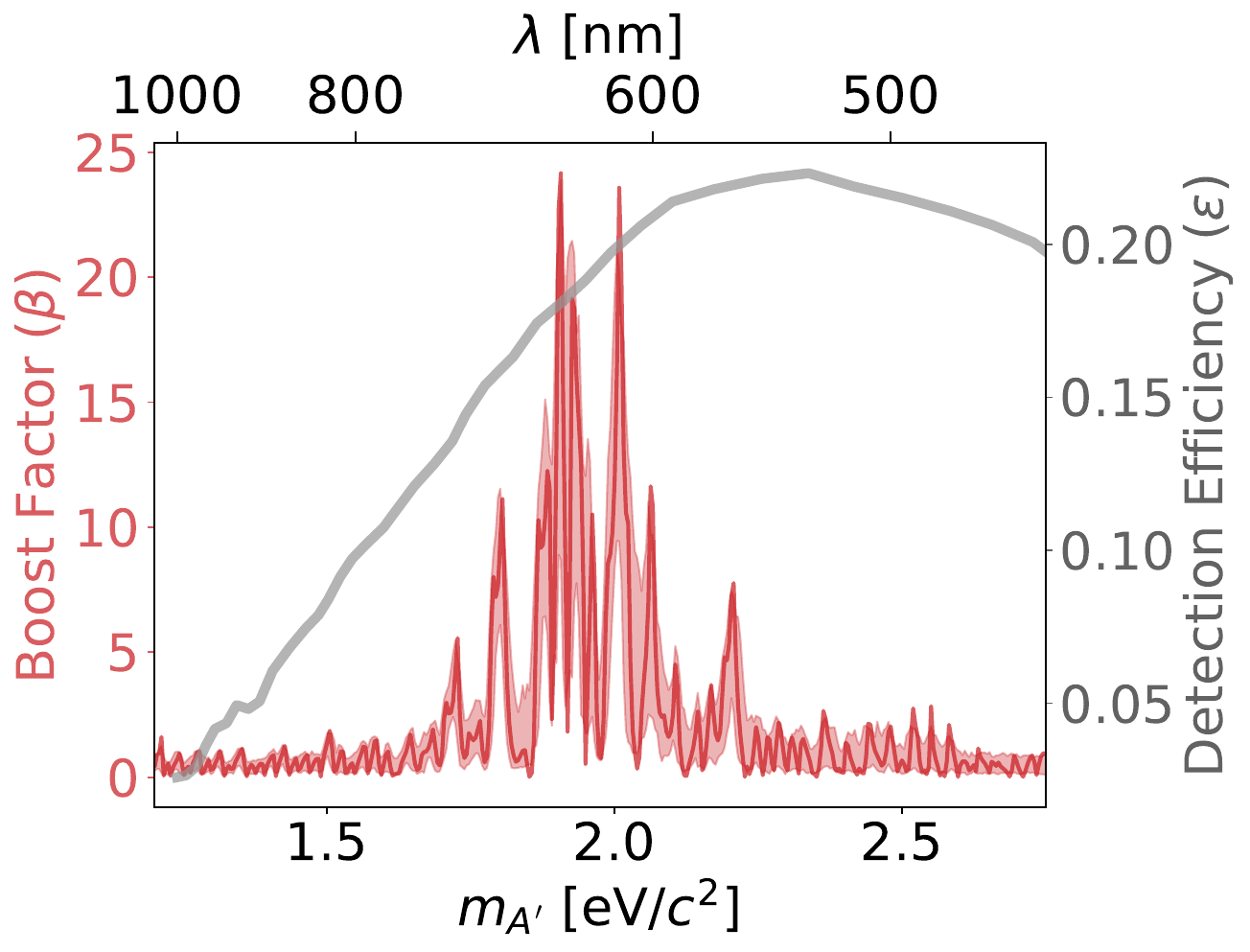}
    \caption{The DPDM conversion boost factor $\beta$ (red) and overall detection efficiency $\epsilon$ (gray) used in Eq.~\ref{eqn:DP_rate}. The boost factor (see Eq.~\ref{eqn:boost}) for a stack with $N = 47$ dielectric pairs is shown as a function of the DM mass $m_{A'}$. The red curve is obtained from theoretical calculations using the layer thicknesses measured during the TEM scan, and the red shaded band indicates the range resulting from the estimated fluctuations of the layer thicknesses. We adopt this TEM-based, as-built profile for the limit calculation as a conservative estimate of the realized prototype performance, while the corresponding idealized maximum boost factor is about 35. The detection efficiency of the converted photons is computed from the calibrated efficiency at 633\,nm together with the relative frequency dependence of the manufacturer CMOS QE curve.
}
    \label{fig:boost}
\end{figure}

\textit{Signal and Background Modeling.---}We define the science run (SR) as data acquired with the dielectric stack installed and the background run (BR) as data acquired with the stack removed, with the optical chain otherwise kept in the same post-calibration configuration. Each exposure contains potential DPDM-converted photons and instrumental noise from the sensor and readout electronics. For DPDM-converted photons, the expected detected count per week, $N_\gamma$, is~\cite{Manenti:2021whp}

\begin{equation}
\label{eqn:DP_rate}
\begin{split}
N_\gamma =
19 &\cdot\frac{\rho}{0.4\,\mathrm{GeV/cm^3}}
\frac{2\,\mathrm{eV}}{m_{A'}}\left(\frac{\kappa}{10^{-12}}\right)^2 \frac{A}{4.9\,\mathrm{cm^2}} \\
& \cdot \epsilon \beta^2\, \langle \cos^2\theta \rangle,
\end{split}
\end{equation}
where $\rho=0.4~\mathrm{GeV/cm^3}$ ($c$=1) is the assumed local DM density, $m_{A'}$ and $\kappa$ are the DPDM mass and kinetic-mixing parameter, and $\langle \cos^2\theta \rangle$ accounts for the DPDM polarization with respect to the detector plane. We average this factor to $2/3$ for a randomly polarized DPDM field~\cite{Caputo:2021eaa}. Here $A$ is the stack area and $\epsilon$ is the overall detection efficiency, including the lens transmission, ROI collection, and effective camera response. The $N$ dielectric pairs enhance the conversion through the boost factor $\beta$, shown in Fig.~\ref{fig:boost}. We compute $\beta$ from the transfer- and source-matrix description of the realized stack configuration, following Ref.~\cite{Millar:2016cjp} and the Supplemental Material~\cite{suppmat}. Its amplitude scales with the pair number $N$ and refractive indices $n_{1,2}$:
\begin{equation} \label{eqn:boost}
\beta \propto N \left( \frac{1}{n_2^2} - \frac{1}{n_1^2} \right).
\end{equation}
In extracting the limit, we use the as-built layer-thickness profile measured by transmission electron microscopy (TEM) rather than the nominal design. The quoted sensitivity therefore corresponds to the realized prototype and is conservative with respect to the idealized stack response.

We model the noise with two components: Gaussian readout noise from charge-to-voltage conversion and digitization, and a Poissonian term describing electrons accumulated during the exposure from dark current and any incident photons. The camera digital number (DN) is modeled as a linear response to the collected charge, with the offset and gain calibrated independently for each pixel. A dedicated short-exposure set of 5000 dark frames with 1\,ms exposure constrains the readout-noise term, for which time-integrated contributions are negligible. We find a median readout noise of 4.45\,DN/frame/pixel, corresponding to 0.806\,$e^-$/frame/pixel. Long-exposure BR data constrain the time-integrated dark charge to $\mathcal{O}(10^{-5})~e^-\,\mathrm{s}^{-1}\,\mathrm{pixel}^{-1}$ and show stable behavior over the full run. Ambient light is suppressed by the lens tube and external enclosure. Dedicated exposures without the ND filters constrain the residual light-leakage rate to approximately $0.2$ photons per pixel per hour, making it negligible after the attenuation of optical-density-12 (OD12) used in the formal SR and BR datasets. The inference is restricted to a retained $60\times60$-pixel ROI chosen to contain most of the DPDM-converted photons and to avoid persistently noisy regions. The resulting per-pixel response model is propagated through the likelihood, preserving few-photon focal-plane information rather than collapsing the data to a single counting variable.

\begin{table}[!tbp]
\caption{\label{tab:calruns} Summary of calibration data sets.}
\vspace{0.1cm}
\centering
\scriptsize
\setlength{\tabcolsep}{2.0pt}
\renewcommand{\arraystretch}{1.08}
\begin{tabular}{@{}lccc@{}}
\toprule
 & \textbf{Cal-I} & \textbf{Cal-II} & \textbf{Cal-III} \\ \midrule
Duration & \shortstack{1\,ms$\times$5k\\$\times$4 sets} & 1\,s$\times$10 & \shortstack{24\,h\\(12\,h on/off)} \\
Focus lens & No & Yes & Yes \\
Attenuation & Tunable & Moderate & $\sim$OD16 \\
Output & gain/offset, $\sigma_{\rm read}$ & focus, geometry & $\epsilon(\omega_0)$, test signal \\
Used in & DN$\rightarrow e^-$ model & template \& MC & Eq.~(\ref{eqn:DP_rate}) norm. \\
\bottomrule
\end{tabular}
\end{table}

\textit{Calibration.---}We perform \textit{in situ} calibration with a 633\,nm He--Ne laser at the stack's maximum-conversion wavelength to (i) define the ROI and optimal focus, (ii) determine the overall detection efficiency $\epsilon$, and (iii) construct the DPDM spatial template and its associated uncertainty. Table~\ref{tab:calruns} summarizes the calibration data sets, and the Supplemental Material~\cite{suppmat} gives the pixel-calibration and optical-simulation details.

\begin{figure}[!tbp]
    \includegraphics[width=\linewidth]{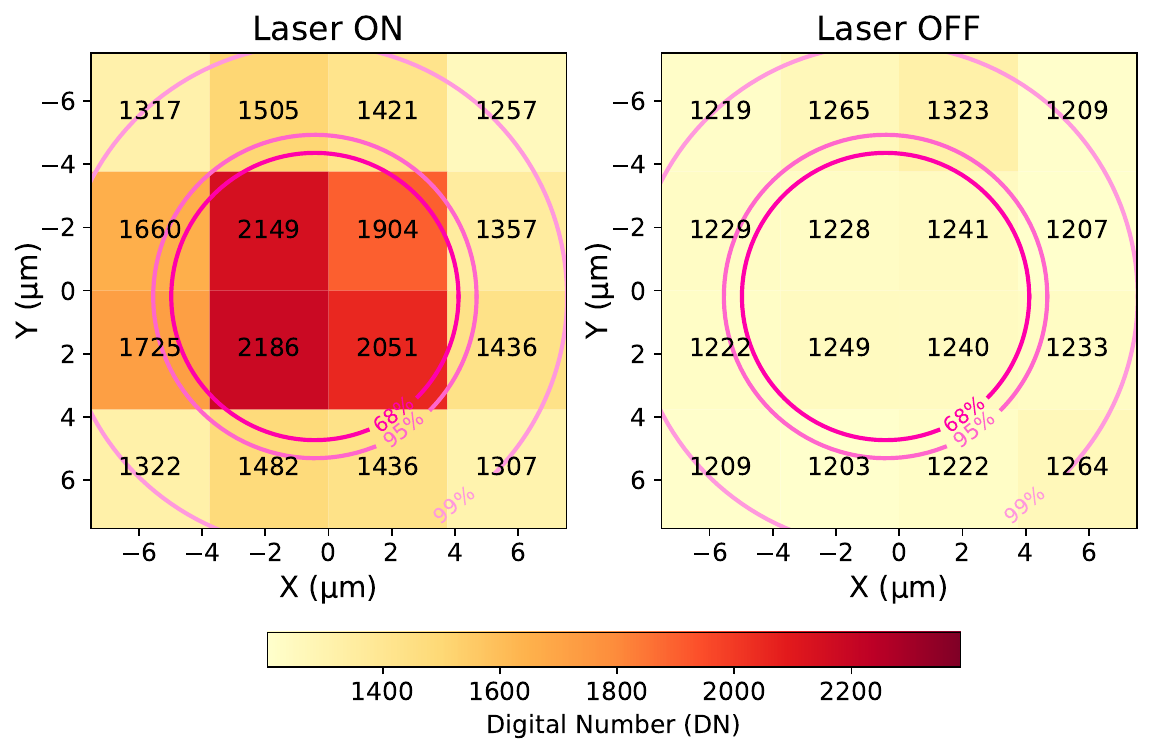}
    \caption{Observed Cal-III patterns in the $4\times4$-pixel ROI, comprising 12 h of attenuated laser-on data (left) and 12 h of laser-off data (right). Pixel values are given in digital numbers. The magenta contours show the expected focal-plane light pattern for DPDM-like illumination, obtained by propagating the calibrated Gaussian mode through the optical simulation.}
    \label{fig:pattern}
\end{figure}

In Cal-I, the laser directly illuminates the CMOS without the stack or lens. We use one dark set and three illuminated sets with different attenuation to determine the per-pixel electronic offset and readout noise from the dark frames and the DN-to-electron gain from the illuminated frames. These parameters define the per-pixel response model that converts DN to collected charge and enters the likelihood. In Cal-II, we steer the collimated laser beam to near-normal incidence on the stack and image it through the focusing lens. We adjust the lens position to optimize the focus and define the ROI while excluding persistently noisy regions. With approximately $8.4\times10^4$ photons contained in the ROI, the resulting light pattern characterizes the optical geometry of the system and helps constrain the expected DPDM spatial template used in the subsequent analysis.
Fig.~\ref{fig:pattern} shows Cal-III data acquired in the few-photon regime, with 12 laser-on and 12 laser-off frames each taken using 1-hour exposures after a total attenuation of $\sim$OD16. The attenuation consists of upstream attenuation before the beam splitter followed by an additional $\sim$OD12 attenuation in the sample arm. A temporary power meter monitors the other arm of the 50:50 beam splitter and measures $15.8$~nW.
This weak-light data anchors the absolute response of the optical chain at $\omega_0$, while the mapping from the calibrated Gaussian laser mode to the DPDM-like near-normal emission mode is obtained from the simulation described below.

To determine the optical collection efficiency, $\eta_{\rm OCE}$, and the DPDM signal template, we supplement the laser calibration with optical simulations. We first tune the simulation to reproduce the observed laser pattern on the CMOS for a collimated Gaussian beam~\cite{Kogelnik:1966AO}. This fit yields the best-focus coordinates in the CMOS $x$--$y$ plane and the longitudinal lens-to-CMOS separation, $z$, and gives a $4\times4$-ROI collection efficiency $\eta_{\rm OCE}^{\rm Gauss}=81.8\%$ for the Gaussian laser beam in Cal-III data.
With the optical geometry fixed by calibration, we generate the DPDM signal template by replacing the Gaussian input with a nearly uniform photon field of small angular divergence incident on the stack interfaces. The broader focal-plane pattern expected for DPDM-like illumination would lead to a small collection efficiency in the compact $4\times4$-pixel region. We therefore use an enlarged $60\times60$-pixel signal ROI, for which the optical collection efficiency is $\eta_{\rm OCE}^{\rm DP}=45.6\%$. This efficiency is included in the signal-amplitude normalization. Combining this DPDM-like collection efficiency with the measured lens transmission and the calibration-anchored camera response, we obtain $\epsilon(\omega_0)=18.8\%$ at $\omega_0$ for the limit extraction. Away from $\omega_0$, we use the manufacturer's quantum-efficiency curve only for the relative wavelength dependence in the vicinity of 633\,nm.
To propagate the uncertainty in the signal pattern, we construct a morphed signal template parameterized by $z$, allow the template to vary over the range defined by the fitted $z$-position uncertainty, and incorporate this morphed template into the inference, as described in the Supplemental Material~\cite{suppmat}. Cal-II and Cal-III are performed immediately before SR1 data-taking. Once we fix the stack and lens positions in Cal-II, the mechanical configuration of the optical chain is not altered during the subsequent SR and BR periods.

\begin{table}[!tbp]
\caption{\label{tab:dataperiod} Summary of data sets used in the analysis.}
\vspace{0.15cm}
\centering
\footnotesize
\setlength{\tabcolsep}{2.5pt}
\begin{tabular}{@{}lccc@{}}
\toprule
 & \textbf{SR1} & \textbf{BR} & \textbf{SR2} \\ \midrule
Configuration & stack-on & stack-off & stack-on \\
Period (2025) & Jul.\,24--Aug.\,20 & Aug.\,20--Sep.\,6 & Sep.\,8--Sep.\,22 \\
Exposure & 572 h & 404 h & 332 h\\
Frame time & 3600 s & 3600 s & 3600 s \\
Purpose & DPDM Search & Noise constraint & DPDM Search  \\
\bottomrule
\end{tabular}
\end{table}

\textit{Data acquisition and analysis.---}Table~\ref{tab:dataperiod} summarizes the science and background exposures used in this work. The first science run (SR1) was acquired with the 47-pair stack installed from July~24 to August~20,~2025, followed by background run (BR) without the stack from August~20 to September~6,~2025. A second science run (SR2) was then collected from September~8 to September~22,~2025. In total, SR1+SR2 contains 904 one-hour exposures, while BR contains 404 one-hour exposures. The CMOS temperature remained stable at $-25.00 \pm 0.06\,^{\circ}\mathrm{C}$ throughout the run.
The final inference used the retained $60\times60$ ROI together with the calibrated spatial template and noise model.

The analysis uses the spatially resolved focal-plane distribution expected from near-normal DPDM conversion together with the calibrated optical response. Rather than reducing the data to a single counting observable, we compare the SR data to a signal model constrained by the laser calibration and optical simulations (see the Supplemental Material~\cite{suppmat}). In this template-based inference, any candidate excess must be consistent with both the total rate and the calibrated focal-plane light pattern.
Before unblinding the final SR1+SR2 search result, we fixed the ROI and morphed signal template using the optical calibration. We determined the per-pixel offset and readout-noise models independently from the short-exposure pixel calibration and incorporated them into the macro-pixel likelihood described in the Supplemental Material~\cite{suppmat}. The likelihood construction, including the treatment of systematic uncertainties, was then specified by these calibration inputs. The expected spatial pattern gives a morphology constraint on any excess while remaining robust against pixel-to-pixel nonuniformities and local background fluctuations.
We propagate the systematic uncertainty associated with the signal model, including the $x$--$y$ alignment and the focal position along $z$, by repeating the inference over the template ensemble and taking the spread in inferred signal normalization as the systematic effect. This uncertainty enters as a contribution proportional to $\sqrt{\mathrm{sig}}$ and is propagated linearly to $\kappa$ (Table~\ref{tab:systematics}).

\begin{figure}[!tbp]
    \includegraphics[width=\linewidth]{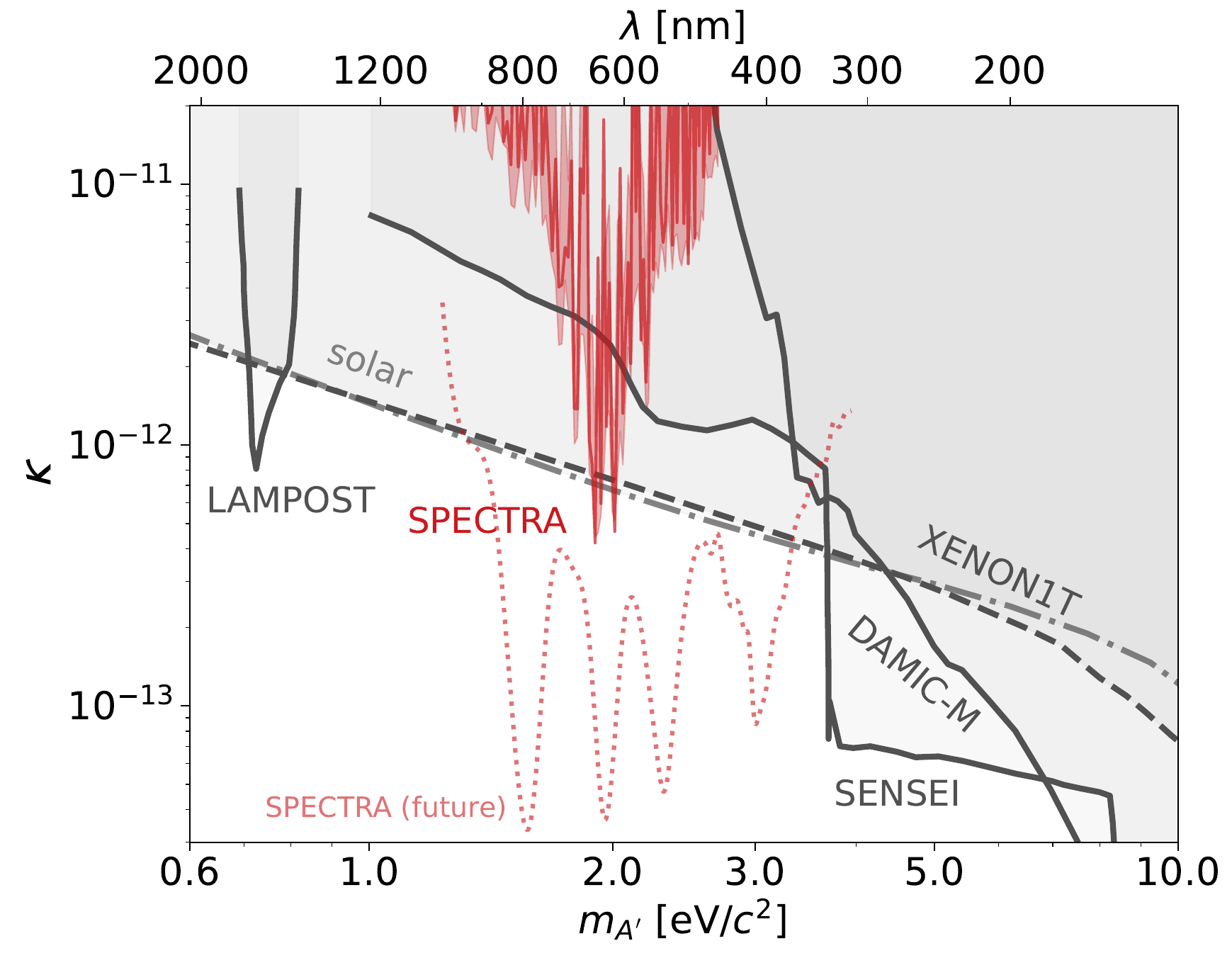}
    \caption{SPECTRA prototype exclusion limit on dark-photon dark matter (DPDM) in the $(m_{A'},\kappa)$ plane at 90\% C.L. Direct laboratory limits on local DPDM from LAMPOST~\cite{PhysRevLett.128.231802}, DAMIC-M~\cite{aguilar2019constraints}, and SENSEI~\cite{PhysRevLett.125.171802} are shown as solid dark curves for comparison. Constraints not relying on local DPDM, including the XENON1T recast of solar dark photons (dashed)~\cite{PhysRevD.102.115022,XENON:2021qze} and stellar energy-loss bounds (gray dash-dotted)~\cite{li2023production} are also shown. The dotted red curve gives the conservative four-stack SPECTRA reach projection described in the text and the Supplemental Material~\cite{suppmat}.}
    \label{fig:limit}
\end{figure}

\textit{Results.---}We observe no statistically significant excess consistent with the expected DPDM template in the combined SR1+SR2 dataset. We therefore derive an exclusion limit on the kinetic-mixing parameter $\kappa$ using a frequentist profile-likelihood analysis calibrated with pseudo-experiments. The resulting 90\% C.L. upper limit is most stringent at $m_{A'}c^2=1.9\,\mathrm{eV}$, where $\kappa < 4.0 \times 10^{-13}$. This corresponds to an improvement in sensitivity by about a factor of two compared with a pure counting analysis ($\kappa < 8.3 \times 10^{-13}$) that neglects the spatial signal pattern in the same ROI. This gain highlights the role of the calibrated focal-plane light pattern as a discriminating observable in the likelihood analysis.
The Supplemental Material~\cite{suppmat} provides the details of the statistical procedure and pseudo-experiment calibration.

\textit{Conclusions and discussion.---}
We report a room-temperature, template-calibrated dielectric-haloscope search for DPDM at optical frequencies. The prototype combines weak-light \textit{in situ} calibration, optical simulation of the DPDM-like near-normal emission profile, and template-based inference in the few-photon regime. In 904 h of stack-on data and 404 h of stack-off control data, we observe no excess and set a localized constraint near $2~\mathrm{eV}$.
The signal normalization is based on the experimentally measured end-to-end optical efficiency and a simulation-derived collection efficiency for the DPDM-like emission mode, providing a conservative estimate of the detected signal yield.
The template-based analysis improves the sensitivity by roughly a factor of two relative to a counting-only approach, demonstrating that the calibrated focal-plane pattern provides important discrimination power.

The same framework can be extended to a multi-stack architecture. Fig.~\ref{fig:limit} shows an illustrative reach projection with four 2-inch, 50-pair stacks, a common CMOS readout with 10~$\mu$m pixel pitch and manufacturer-provided QE, an eight-fold reduction in readout noise, and three months each of stack-on and stack-off exposure. The stack peak frequencies are chosen to match common calibration-laser wavelengths, preserving the same in situ calibration and focal-plane-template strategy. The reach curve uses a conservative stack-response model with representative layer-thickness fluctuations and is based only on the total signal-yield sensitivity of the four modules. Under these assumptions, the projected kinetic-mixing reach improves by one to two orders of magnitude relative to the present prototype.

The multi-stack layout also adds spectral leverage because a dark-photon signal would produce a characteristic pattern of relative amplitudes across the differently tuned stacks. In the coverage comparison described in the Supplemental Material~\cite{suppmat}, four laser-aligned stacks outperform equal-area single-stack references over roughly three quarters of the 1.45--3.10~eV band, with median discovery-significance gains of about 2.6--2.7. If a candidate excess appears, the same relative-amplitude pattern could enable cross-stack mass localization, test cross-stack consistency, and guide follow-up measurements. More broadly, replicated, calibrated room-temperature stack modules coupled to a common imaging readout could provide a practical route for extending template-calibrated eV-scale DPDM searches.

\textit{Acknowledgments.---}We thank Youxi Li, Xinyu Chen, Zhenhao Liang, and Zhu Liu for their assistance with the optical setup. We thank Fanglin Bao for helpful discussions. We thank Laura Manenti and collaborators for making their boost-factor calculation code publicly available. We are also grateful to the Instrumentation and Service Center for Physical Sciences (ISCPS) at Westlake University for providing the TEM measurements. SL acknowledges support from the National Natural Science Foundation of China and the Start-up Funding of Westlake University.

\bibliography{dp_paper}

\clearpage
\onecolumngrid
\suppressfloats[t]
% ---------- Supplemental numbering ----------
\setcounter{section}{0}
\renewcommand{\thesection}{S\arabic{section}}
\setcounter{equation}{0}
\renewcommand{\theequation}{S\arabic{equation}}
\setcounter{figure}{0}
\renewcommand{\thefigure}{S\arabic{figure}}
\setcounter{table}{0}
\renewcommand{\thetable}{S\arabic{table}}
% Hyperref anchors, if hyperref is loaded
\makeatletter
\@ifpackageloaded{hyperref}{
  \renewcommand{\theHsection}{suppsection.\arabic{section}}
  \renewcommand{\theHequation}{suppequation.\arabic{equation}}
  \renewcommand{\theHfigure}{suppfigure.\arabic{figure}}
  \renewcommand{\theHtable}{supptable.\arabic{table}}
}{}
\makeatother
% A robust supplement section command for PRL/REVTeX
\newcommand{\suppsection}[2]{%
  \refstepcounter{section}%
  \section*{\thesection\quad #1}%
  \label{#2}%
}

\begin{center}
  {\large{\textbf{\textit{Supplemental Material:\\ Searching for Dark Photons with a Room-Temperature Dielectric Haloscope}}}}
\end{center}

\suppsection{Stack fabrication and TEM characterization}{app:stack}
\paragraph{\underline{Stack fabrication.}} We fabricated the dielectric conversion medium on a double-side-polished 1-inch (25.4\,mm) sapphire substrate with thickness 430\,$\mu$m and surface roughness $R_a < 0.2$\,nm. We first cleaned the substrate with oxygen plasma (PlasmaEtch PE-25) to remove organic contaminants and improve film adhesion. We then transferred it to a Temescal BJD-200 electron-beam evaporator and applied 120~s of argon-plasma bombardment to remove residual surface contaminants, including native oxides. We deposited a 105-nm Ag layer and capped it with 23~nm of Ta$_2$O$_5$, with both materials evaporated at 5~\AA/s.

After evaporation, we sonicated the structure in deionized water for 10\,min, dried it with 99.99\% nitrogen, mounted it in an ion-beam sputtering deposition (IBSD) reticle system from Angstrom Engineering, cleaned it again with argon plasma, and deposited 47 TiO$_2$/SiO$_2$ dielectric pairs with refractive indices $n_1 \sim 2.34$ and $n_2 \sim 1.43$. In this design, each dielectric layer satisfies $n d = 316$\,nm, corresponding to $d=135.3$\,nm for TiO$_2$ and $d=221.5$\,nm for SiO$_2$. This choice optimizes dark-photon-to-photon conversion near 633\,nm. The Ag layer acts as a mirror that enhances the boost factor and reflects converted photons toward the camera. The thin Ta$_2$O$_5$ layer protects the Ag film during transfer and deposition and improves adhesion to the dielectric pairs.

\begin{figure}[t]
    \includegraphics[width=0.3\linewidth]{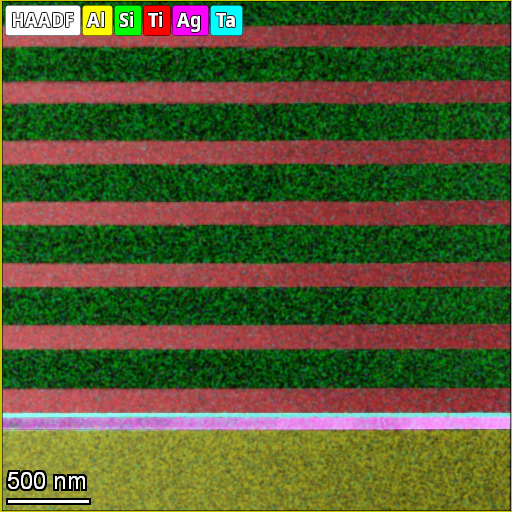}
    \caption{Transmission electron microscopy (TEM) image of the stack near the substrate. The scale bar appears in the lower-left corner, and the colors denote different elements. From bottom to top, the layers correspond to the sapphire (Al$_2$O$_3$) substrate, the Ag mirror, the Ta$_2$O$_5$ layer, and the alternating TiO$_2$ and SiO$_2$ dielectric layers.}
    \label{fig:TEM}
\end{figure}

\paragraph{\underline{Characterization.}} We characterized the stack with reflectivity measurements and transfer-matrix method (TMM) calculations to verify the layer growth and thickness uniformity. For the reflectivity measurement, we coupled an HL-2000\_HP white-light source through an optical fiber to a USB4000-VIS-NIR spectrometer (Ocean Insight). We compared the measured spectrum with the TMM prediction to validate the Ag mirror fabricated in the electron-beam evaporator. At the wavelength of interest, 633\,nm, the measured reflectivity is 94\%, in agreement with the calculation.
After the science run, we characterized the dielectric stack using transmission electron microscopy (TEM) and extracted the thickness mean and variation of each layer. Fig.~\ref{fig:TEM} shows a zoomed-in TEM image. We then compute the nominal boost factor~\cite{Millar:2016cjp} using the TEM-measured thicknesses, as shown by the solid curve in Fig.~\ref{fig:boost}. To propagate the thickness uncertainties, we generate $10^4$ stack realizations by sampling the layer thicknesses according to the TEM-measured variations. The red shaded band in Fig.~\ref{fig:boost} shows the resulting 1-$\sigma$ interval of the boost factor at each mass point.

\suppsection{Pixel calibration, pattern calibration, and simulation}{app:calib}
We calibrate the per-pixel offset, readout noise, and gain of each CMOS pixel with short-exposure dark frames and controlled laser illumination at several intensity settings. We then simulate Gaussian and nearly uniform beams through the focusing lens to determine the three-dimensional focal position and the corresponding intensity distribution on the sensor. These steps define the calibration framework and the signal-pattern inputs used in the subsequent analysis.

\begin{figure}[t]
    \includegraphics[width=0.65\linewidth]{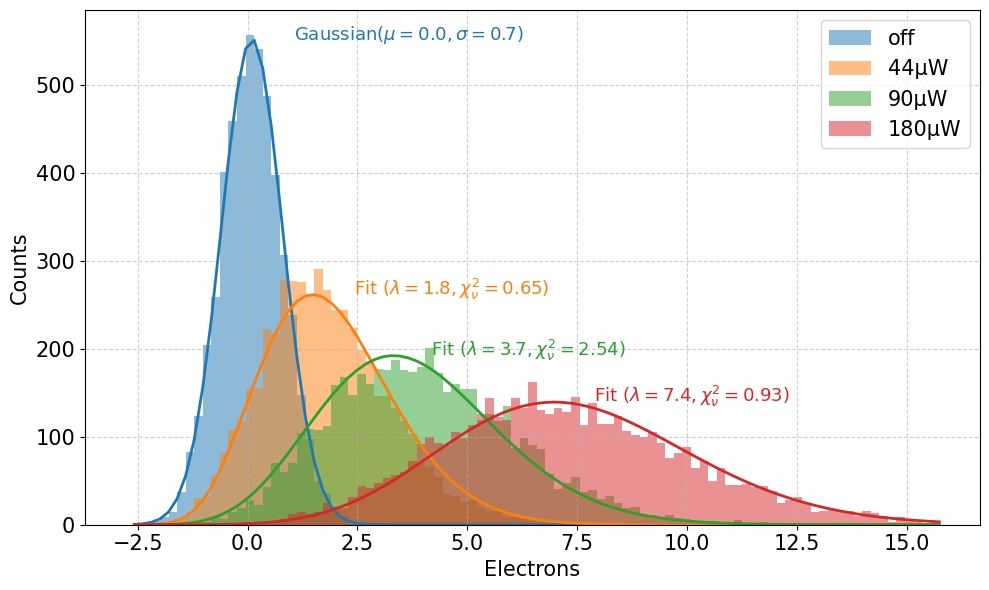}
    \caption{Calibration data from one representative pixel at proportional laser intensities. The blue, orange, green, and red regions correspond to the laser-off, $1\times$, $2\times$, and $4\times$ settings, respectively, under defocused short-exposure conditions. For each fixed intensity, we record 5000 frames. The fit shows that the single-pixel response model in Eq.~\ref{eqn:pixel} describes the data under all tested laser intensities. The legend gives the expected laser power before the ND filter.}
    \label{fig:offset}
\end{figure}

\paragraph{\underline{Pixel calibration.}} We extract the per-pixel offset and readout noise from the short-exposure dark frames described in the main text. To measure the gain, we illuminate the CMOS with an external laser attenuated by a neutral-density (ND) filter. During this calibration, the CMOS is held at the same $-25^{\circ}\mathrm{C}$ temperature and operated with a 1-ms exposure time. We monitor the laser power in real time with a beam splitter and a power meter and set the incident power to $44~\mu\mathrm{W}$, $90~\mu\mathrm{W}$, and $180~\mu\mathrm{W}$ before the fixed ND filter. At each setting, we record 5000 frames. Under these conditions, the digital-number (DN) model for a single pixel is

\begin{equation}
DN = \mathcal{N}(N_0, \sigma_\text{read}) + g\cdot \textrm{Pois}(\lambda),
\label{eqn:pixel}
\end{equation}
where $\lambda$ is the expected photon count for a given laser intensity, $N_0$ is the baseline, and $\sigma_\text{read}$ is the readout noise. As shown in Fig.~\ref{fig:offset}, we fit all intensity settings simultaneously with relative amplitudes 0, $1\times$, $2\times$, and $4\times$, and extract the best-fit gain $g$ and offset $N_0$ for each pixel. We then propagate their values to the inference model. For illustration, the selected ROI pixels show distributions of the parameters with $N_0 = 101.3^{+0.8}_{-0.4}~\mathrm{DN}$, $\sigma_\text{read}=4.0^{+1.1}_{-0.4}~\mathrm{DN}$, and $g=5.1^{+0.9}_{-0.2}~\mathrm{DN}/e^-$, corresponding to the 16th, 50th, and 84th percentiles across the ROI.

\paragraph{\underline{Pattern calibration.}} We determine the three-dimensional focal position by combining the laser-calibration data with the optical simulations. We extract the transverse coordinates $x$ and $y$ from calibration images taken with low optical-density filtering, where the laser intensity is high enough to produce a clean pattern. A centroid reconstruction gives the initial estimate of the focal position in the transverse plane.

We model the laser spot after the lens as a two-dimensional Gaussian on the CMOS, corresponding to the Fourier-transformed beam profile. Fitting this distribution along the $x$ and $y$ directions yields the final transverse position. The resulting $4\times4$ ROI is $x\in[439,442]$ and $y\in[657,660]$ pixels, with fitted absolute center coordinates $(x_c, y_c)=(\,440.40, 658.55)$ pixels.

\begin{figure}[t]
    \includegraphics[width=0.8\linewidth]{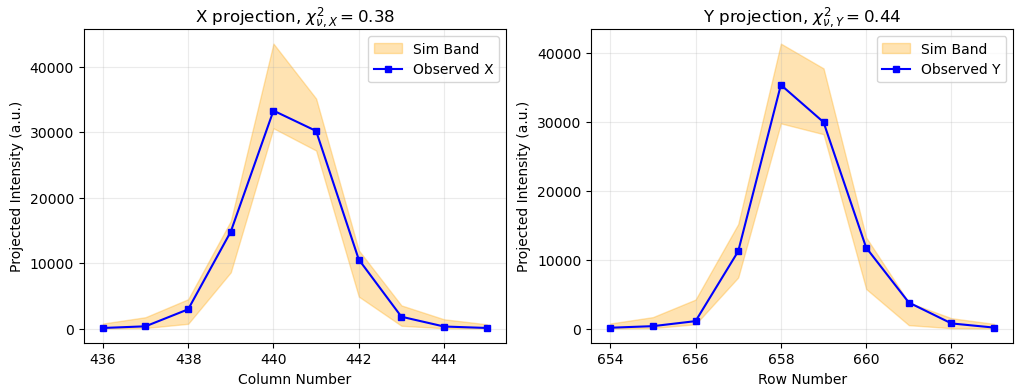}
    \caption{X and Y projections of the signal pattern during laser calibration (Cal-II). The blue points represent the measured calibration data. The orange band indicates the simulated intensity profiles obtained by varying the longitudinal position $z$ around the best-fit focus and taking the resulting envelope as the uncertainty band of the expected pattern. The quoted $\chi^2_{\nu}$ values are computed by comparing the observed profile to the best-fit simulated pattern, with uncertainties given by the width of the simulation pattern band.}
    \label{fig:opisim}
\end{figure}

The longitudinal coordinate $z$, defined as the lens-to-CMOS separation, is obtained by comparing the measured calibration pattern with simulated patterns at different $z$ values and identifying the best-fit focus by minimizing the Poisson deviance. The resulting best-fit value is $z=77.42^{+0.06}_{-0.30}~\mathrm{mm}$. To account for the associated pattern uncertainty (cf. Table~\ref{tab:systematics}), we repeat the optical simulations for representative $z$ values within this range and incorporate the resulting pattern variations into the statistical analysis, as shown in Fig.~\ref{fig:opisim}.

\paragraph{\underline{Signal pattern simulation.}}
We simulate the focal-plane signal pattern using ray-tracing-based geometric optics at $\lambda=633~\mathrm{nm}$ under ambient conditions of $23\,^{\circ}\mathrm{C}$ and $1.0~\mathrm{atm}$. The ray-tracing calculation is implemented in the open-source Python optical-design package
Optiland~\cite{Optiland}, while representative configurations were qualitatively cross-checked
with an independent commercial optical-design package~\cite{ZemaxOpticStudio}. We first propagate a Gaussian calibration beam through the focusing lens and compare the simulated focal spot with the CMOS calibration data, thereby determining the effective focus position and lens response in the realized setup. The calibrated optical geometry is then used to simulate DPDM-induced emission as a nearly spatially uniform input field with a small angular divergence. The divergence angles are randomly sampled from the Maxwellian velocity distribution of the dark-photon field~\cite{Baryakhtar2018}. The resulting templates provide both the expected focal-plane pattern and the overall light-collection efficiency. The $60\times60$-pixel ROI collects nearly 100\% of the Gaussian calibration beam, but only 44\%--55\% of the DPDM-like illumination across the signal templates. This reduction is driven mainly by the angular spread of the DPDM-like illumination, which causes a significant fraction of rays to fall outside the finite focal-plane ROI.

\begin{figure*}[t]
    \centering
    \includegraphics[width=0.95\textwidth]{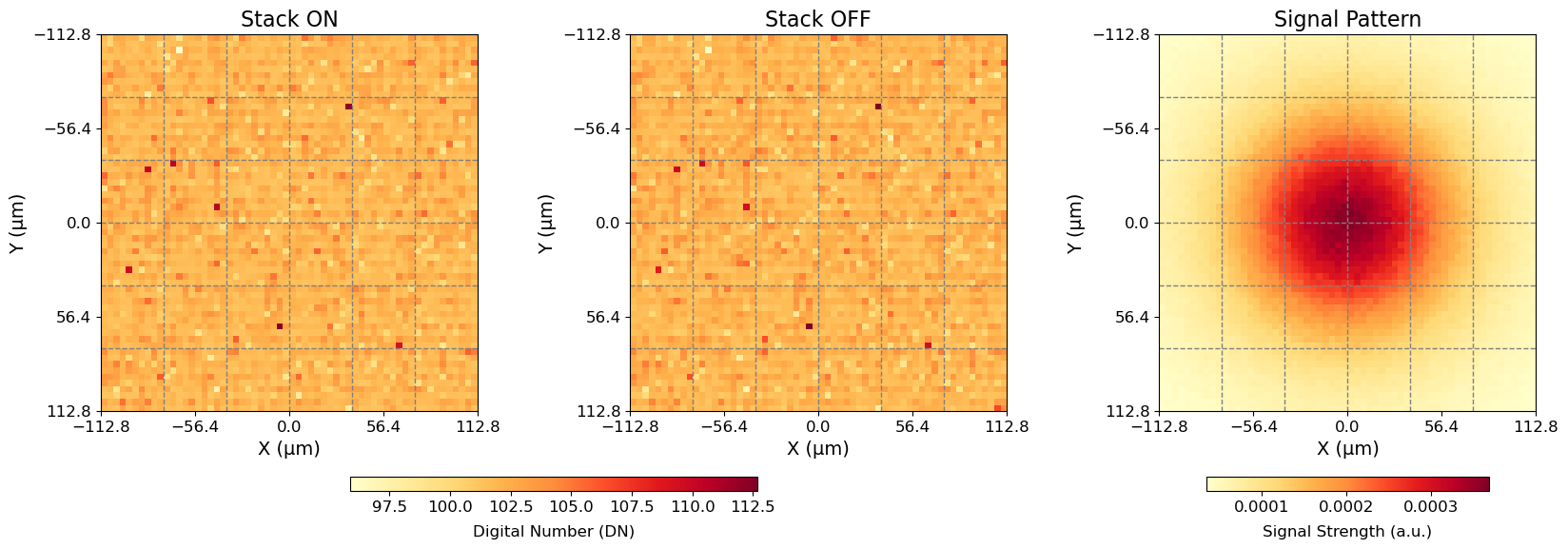}
    \caption{
    Spatial distributions in the $60\times60$-pixel ROI. The left and middle panels show the total DN per hour measured during the SR with the dielectric stack installed (904 h) and the BR with the stack removed (404 h), respectively. These two panels share the same color scale, allowing a direct pixel-by-pixel comparison. The right panel shows the simulated signal pattern used in the likelihood analysis, displayed with an independent color scale. The dashed lines indicate the boundaries of the $6\times6$ macro-pixels.
    }
    \label{fig:result}
\end{figure*}

\suppsection{Statistical inference}{app:stat}
\paragraph{\underline{Frequentist profile-likelihood inference.}}
We quantify a potential DPDM signal above background with a frequentist profile-likelihood-ratio (PLR) analysis. The inputs are two stacked \(60\times 60\) ROI images,
\(DN^{\mathrm{on}}\) and \(DN^{\mathrm{off}}\), corresponding to the
stack-on SR and stack-off BR control configurations,
respectively. As shown in Fig.~\ref{fig:result}, to make the inference robust and
computationally tractable, we rebin each \(60\times 60\) ROI into a
\(6\times 6\) array of macro-pixels, where each macro-pixel aggregates
the corresponding sub-region of the original ROI. The inference is then
performed on these coarse-grained \(6\times 6\) ROI summaries rather
than the full pixel-level images to reduce computational cost and statistical fluctuations.

\paragraph{\underline{Macro-pixel construction.}}
The inference is performed using \(36\) macro-pixels constructed from the original 3600 camera pixels. 
We denote the original pixel index by \(i\), and the macro-pixel index by 
\(m=1,\ldots,36\). Each macro-pixel contains a set of original pixels 
\(\mathcal{P}_m\). The gain \(g_i\), electronic offset \(N_{0,i}\), and 
single-frame readout noise \(\sigma_{\mathrm{read},i}\) are defined at the 
original pixel level.

The position-dependent signal template is defined at the pixel level as 
\(p_i(z)\), normalized over the active ROI according to
\begin{equation}
\sum_i p_i(z) = 1 .
\end{equation}
The corresponding macro-pixel signal template is
\begin{equation}
p_m(z)
=
\sum_{i\in\mathcal{P}_m} p_i(z) .
\end{equation}

The total signal strength is described by a single position-independent parameter \(S\), while the relative signal efficiency is allowed to vary with detector position through \(\epsilon(z)\). The detector position \(z\) is treated as a nuisance parameter, and both \(p_m(z)\) and \(\epsilon(z)\) are continuously morphed within the calibration-derived allowed
range.
The nominal background template \(B_m\) is determined from the stack-off data. A common normalization factor \(\alpha\) is introduced as a shared nuisance parameter in the stack-on and stack-off likelihoods, allowing the overall background rate to vary while preserving the relative spatial shape of \(B_m\).\par
We also include a fixed, sideband-measured mean DN-offset correction between the stack-on and stack-off samples. Let \(\mathcal{S}_{\rm side}\) denote the set of sideband pixels outside the signal ROI, where no dark-photon signal is expected. For each run configuration \(r\in\{\mathrm{on},\mathrm{off}\}\), we define the pedestal-subtracted per-frame sideband mean as
\[
\overline{d}^{\,r}_{\rm side}
=
\frac{1}{|\mathcal{S}_{\rm side}|}
\sum_{i\in\mathcal{S}_{\rm side}}
\frac{1}{F_r}
\sum_{f=1}^{F_r}
\left(\mathrm{DN}^{r}_{i,f}-N_{0,i}\right).
\]
The fixed DN-offset estimator is then
\[
\hat{\delta}_{\mathrm{DN}}
=
\overline{d}^{\,\mathrm{on}}_{\rm side}
-
\overline{d}^{\,\mathrm{off}}_{\rm side}
=
-3.5\times10^{-2}\ \mathrm{DN\,pixel^{-1}\,frame^{-1}} .
\]
The magnitude of this shift is extremely small, corresponding to only about \(0.8\%\) of the median per-frame readout-noise scale. Nevertheless, we keep it as a fixed correction in the stacked likelihood to remove a possible coherent mean offset between the stack-on and stack-off samples, which does not alter the spatial shape of \(B_m\). The expected macro-pixel means are then
\begin{align}
&\mu_{\mathrm{on},m}(S,z,\alpha)
=
S\,\epsilon(z)\,p_m(z)
+
\frac{F_{\mathrm{on}}}{F_{\mathrm{off}}}\,\alpha B_m
+
F_{\mathrm{on}}\,\hat{\delta}_{\mathrm{DN}}
\sum_{i\in\mathcal{P}_m}\frac{1}{g_i},
\\
&\mu_{\mathrm{off},m}(\alpha)
=
\alpha B_m .
\end{align}
Here \(F_{\mathrm{on}}\) and \(F_{\mathrm{off}}\) are the numbers of stacked on and off frames. Since the stack-off sideband has a slightly larger mean DN value, the estimator gives \(\hat{\delta}_{\mathrm{DN}}<0\), yielding a small negative correction to the stack-on expectation relative to the stack-off-derived background template.

\paragraph{\underline{Likelihood after DN-to-charge conversion.}}
The stacked DN sums are converted to electron-equivalent counts at the pixel 
level and then summed into macro-pixels:
\begin{align}
e_{\mathrm{on},m}
&=
\sum_{i\in\mathcal{P}_m}
\frac{
DN^{\mathrm{on}}_i
-
F_{\mathrm{on}}N_{0,i}
}{g_i},
\\
e_{\mathrm{off},m}
&=
\sum_{i\in\mathcal{P}_m}
\frac{
DN^{\mathrm{off}}_i
-
F_{\mathrm{off}}N_{0,i}
}{g_i}.
\end{align}
The corresponding macro-pixel readout-noise variances are
\begin{align}
\sigma^2_{\mathrm{on},m}
&=
F_{\mathrm{on}}
\sum_{i\in\mathcal{P}_m}
\sigma^2_{\mathrm{read},i},
\\
\sigma^2_{\mathrm{off},m}
&=
F_{\mathrm{off}}
\sum_{i\in\mathcal{P}_m}
\sigma^2_{\mathrm{read},i}.
\end{align}

Each macro-pixel measurement is modeled as a Poisson-distributed number of 
collected electrons convolved with Gaussian readout noise:
\begin{align}
\mathcal{L}_m(S,z,\alpha)
&=
\sum_{\lambda_{\mathrm{on}}=0}^{\infty}
\operatorname{Pois}
\!\left(
\lambda_{\mathrm{on}}
\mid
\mu_{\mathrm{on},m}
\right)
\mathcal{N}
\!\left(
e_{\mathrm{on},m}
\mid
\lambda_{\mathrm{on}},
\sigma^2_{\mathrm{on},m}
\right)
\nonumber
\\
&\quad \times
\sum_{\lambda_{\mathrm{off}}=0}^{\infty}
\operatorname{Pois}
\!\left(
\lambda_{\mathrm{off}}
\mid
\mu_{\mathrm{off},m}
\right)
\mathcal{N}
\!\left(
e_{\mathrm{off},m}
\mid
\lambda_{\mathrm{off}},
\sigma^2_{\mathrm{off},m}
\right).
\end{align}
The full likelihood is the product over the \(36\) macro-pixels:
\begin{equation}
\mathcal{L}(S,z,\alpha)
=
\prod_{m=1}^{36}
\mathcal{L}_m(S,z,\alpha).
\end{equation}

The parameter of interest is the total signal strength \(S\). The detector 
position \(z\) and the global background scale \(\alpha\) are treated as 
nuisance parameters and are profiled in the construction of the 
profile-likelihood test statistic.

\begin{figure}[t]
    \includegraphics[width=0.65\linewidth]{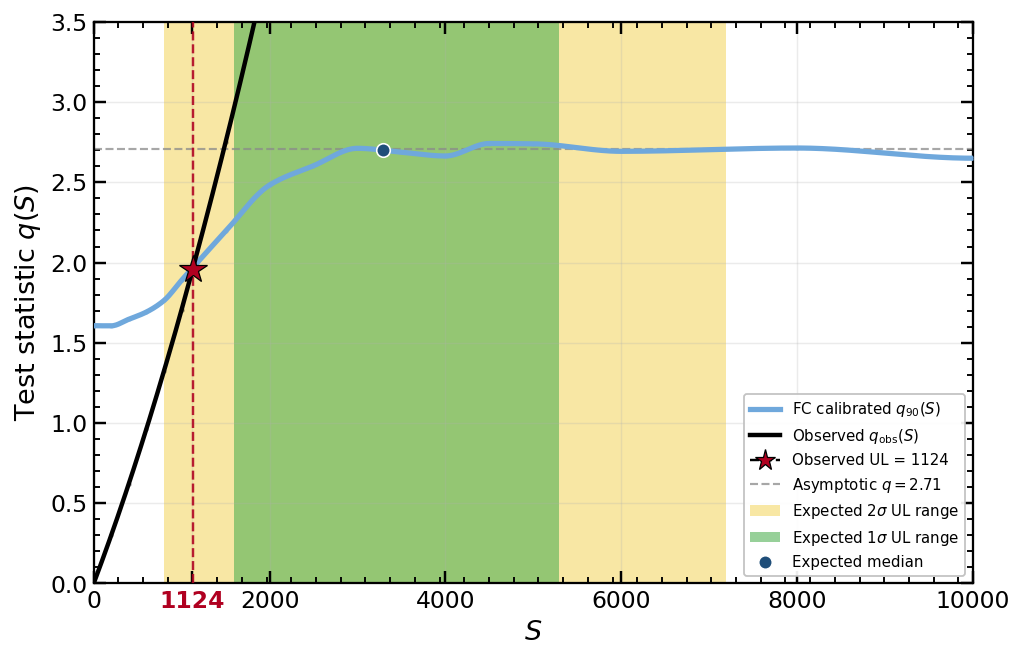}
    \caption{Feldman--Cousins calibrated upper-limit (UL) construction for the signal strength \(S\). The blue curve shows the calibrated 90\% critical value \(q_{90}(S)\), obtained from \(2\times10^{4}\) toy Monte Carlo pseudo-experiments at each tested \(S\). This calibration is used instead of the asymptotic \(q=2.71\) threshold, shown by the gray dashed line, because the finite-sample profiled likelihood with nuisance parameters and the physical boundary \(S\geq0\) does not necessarily follow the asymptotic \(\chi^{2}\) approximation. The green and yellow bands show the central 68\% and 95\% expected upper-limit ranges under the background-only hypothesis using \(5\times10^{4}\) toy Monte Carlo pseudo-experiments, with the blue marker indicating the median expected limit. The observed test statistic \(q_{\rm obs}(S)\), shown in black, crosses the calibrated critical curve at \(S=1124\), marked by the red marker and dashed line, defining the observed 90\% confidence-level upper limit. }
    \label{fig:inference}
\end{figure}

\paragraph{\underline{Profile likelihood ratio.}}
We take the total signal strength \(S\) as the parameter of interest. The 
detector position \(z\), which controls the morphed signal template and 
relative efficiency, and the global background scale \(\alpha\) are treated 
as nuisance parameters. For each fixed value of \(S\), the likelihood is 
profiled over \(z\) and \(\alpha\). The profile likelihood ratio is
\begin{equation}
\lambda(S)
=
\frac{
\mathcal{L}\!\left(S,\hat{\hat{z}}(S),\hat{\hat{\alpha}}(S)\right)
}{
\mathcal{L}\!\left(\hat{S},\hat{z},\hat{\alpha}\right)
},
\end{equation}
where \(\hat{\hat{z}}(S)\) and \(\hat{\hat{\alpha}}(S)\) are the conditional 
maximum-likelihood estimates at fixed \(S\), while 
\((\hat{S},\hat{z},\hat{\alpha})\) are the unconditional maximum-likelihood 
estimates. The fixed macro-pixel background inputs \(B_m\) and the sideband-corrected DN-offset \(\hat{\delta}_{\mathrm{DN}}\) are kept fixed in this profiling procedure.

\paragraph{Test statistic for interval construction.}
To construct confidence intervals for the signal strength \(S\), we use the
profile-likelihood-ratio test statistic
\begin{equation}
q(S) = -2\ln\lambda(S),
\end{equation}
where \(\lambda(S)\) is the profile likelihood ratio evaluated at the tested
signal strength \(S\).

Because the event counts are low and the likelihood is generally non-Gaussian,
we do not rely on the asymptotic \(\chi^{2}\) approximation. Instead, for each
tested signal strength \(S\), we determine the sampling distribution of
\(q(S)\) using pseudo-experiments generated under the hypothesis \(S\). The
Feldman--Cousins critical value \(q_{90}(S)\) is defined by
\begin{equation}
P\!\left(q(S) \le q_{90}(S) \mid S\right) = 0.90 .
\end{equation}
This construction gives the 90\% acceptance region for each tested value of
\(S\) and preserves frequentist coverage in the low-count regime.

\paragraph{Confidence interval and observed upper limit.}
For the observed data, the test statistic as a function of signal strength is
denoted by \(q_{\rm obs}(S)\). The 90\% confidence interval is the set of signal
strengths satisfying
\begin{equation}
q_{\rm obs}(S) \le q_{90}(S).
\end{equation}
The observed one-sided upper limit, \(S_{90}^{\rm obs}\), is defined as the
upper edge of this interval. Equivalently, it is given by the crossing point
\begin{equation}
q_{\rm obs}\!\left(S_{90}^{\rm obs}\right)
=
q_{90}\!\left(S_{90}^{\rm obs}\right).
\end{equation}
In the present analysis this crossing occurs at \(S_{90}^{\rm obs}=1124\), as
shown by the red marker in Fig.~\ref{fig:inference}.

\paragraph{Discovery test and significance.}
For discovery, we use a separate one-sided test of the background-only hypothesis. The discovery test statistic is
\begin{equation}
q_0 =
\begin{cases}
-2\ln \lambda(0), & \hat S \ge 0,\\[4pt]
0, & \hat S < 0,
\end{cases}
\end{equation}
which is nonzero only for upward fluctuations in the fitted signal strength. Its sampling distribution under the background-only hypothesis is calibrated with pseudo-experiments rather than asymptotic approximations. For an observed dataset, the background-only $p$-value is
\begin{equation}
p_0 = P\!\left(q_0 \ge q_0^{\rm obs} \mid S=0\right),
\end{equation}
and the corresponding Gaussian-equivalent significance is defined by
\begin{equation}
Z = \Phi^{-1}(1-p_0),
\end{equation}
where $\Phi$ is the cumulative distribution function of the standard normal distribution.

\paragraph{\underline{Discovery threshold and discovery sensitivity.}}
For discovery-oriented benchmarks, we choose a target significance threshold $Z_\ast$ (e.g., $Z_\ast=3$ or $5$) and determine the corresponding critical value $q_{0,\rm crit}$ from background-only pseudo-experiments according to
\begin{equation}
P\!\left(q_0 \ge q_{0,\rm crit} \mid S=0\right)=1-\Phi(Z_\ast).
\end{equation}
The median discovery sensitivity is then defined as the smallest injected signal strength for which the median value of $q_0$ exceeds $q_{0,\rm crit}$. This definition keeps the discovery metric consistent with the low-count, non-Gaussian regime of the present analysis.

\suppsection{Data analysis}{app:postcheck}
\paragraph{\underline{Analysis pipeline.}}
After data taking, we retained 1308 frames in total: 904 from SR and 404 from BR. Because the analysis targets the signal accumulated in each pixel, we stack the SR and BR frames separately and exposure-normalize them for display, forming two $60\times60$ DN-per-hour matrices shown in the left and middle panels of Fig.~\ref{fig:result}. The right panel shows the mean value of the simulated spatial pattern expected for the signal, before Poisson fluctuations from photon counting. We then apply the per-pixel calibration, including baseline offsets and gains, to convert the stacked DN values to electron-equivalent counts. The BR dataset constrains the background contribution in the SR frames, while the simulated pattern specifies the expected pixel-by-pixel signal distribution. The likelihood described in Sec.~\ref{app:stat} combines these components to derive a 90\% confidence-level upper limit on the signal strength.

\begin{figure*}[t]
    \includegraphics[width=0.9\linewidth]{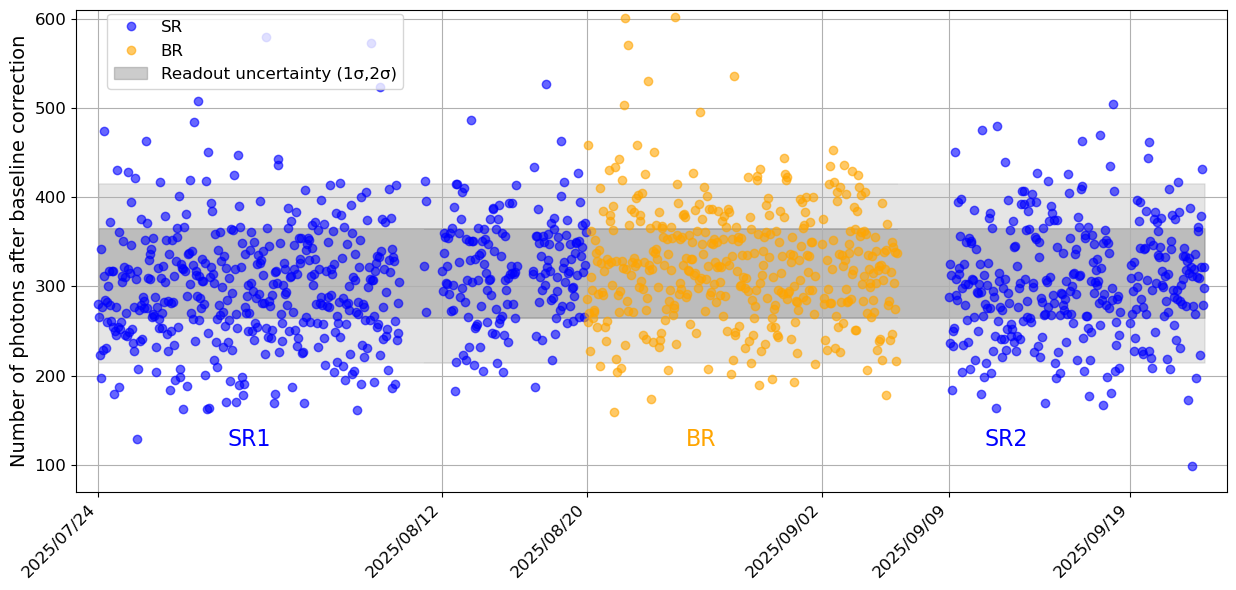}
    \caption{Time evolution of the total signal in the ROI for each frame, baseline-corrected by pixel and converted to electron-equivalent counts. The blue points show Science Run data with the stack installed, and the yellow points show Background Run data with the stack removed. The gray band denotes the fluctuation expected from the readout noise, centered on the mean signal across all frames.
    }
    \label{fig:result_time}
\end{figure*}

Fig.~\ref{fig:result_time} shows the time evolution of the baseline-corrected sum over the 3600-pixel ROI. The per-frame totals remain broadly stable throughout the full data-taking period, with no visible drift or abrupt change across SR1, BR, and SR2. Most frames fluctuate within the readout-uncertainty band. The strong overlap between the projected SR and BR distributions is consistent with stable temperature control and stable detector operating conditions throughout the measurement.

\paragraph{\underline{Error estimation.}}

We determined the laser-spot position on the CMOS with a two-dimensional image analysis. We first subtracted the background and clipped the images to non-negative values. We then used a weighted centroid to obtain the initial spot center $(x_0, y_0)$ and the second-moment widths $(\sigma_x, \sigma_y)$. Near the centroid, we fit the $x$ and $y$ projections with baseline-corrected Gaussian functions and used the covariance matrix of the fit parameters to estimate the uncertainties on the center position and widths.

The optimal focus coordinate $z$ between the lens and the CMOS is obtained by comparing the measured spot distribution with optical simulations at different focus positions (Sec.~\ref{app:calib}). For each candidate distance, we normalized the simulated intensity to the total observed intensity, aligned the simulated pattern with the experimental data, and interpolated it onto the experimental coordinates. We quantified the agreement using the Poisson deviance and identified its minimum as the best-fit focus. Starting from this best-fit value, we progressively enlarged the $z$ interval in both directions until the observed distribution was fully enclosed by the resulting pattern-variation band, as shown in Fig.~\ref{fig:opisim}. The selected interval approximately corresponds to the points at which the Poisson deviance increases by $0.05$ above its minimum and is treated as a systematic range.

We estimated the per-pixel offset and gain from 5000 frames of 1-ms exposures obtained in Cal-I. As described in Sec.~\ref{app:calib}, we take the mean of the dark-field short exposures as the pixel offset, and its uncertainty is given by the Gaussian width divided by the number of exposure frames. Using these measured quantities, we fit the pixel DN sequence with a joint Poisson--Gaussian model. Maximum-likelihood estimation yields the pixel gain and the incident photon number $\lambda$, and the fit covariance provides the gain uncertainty.

For the detector-side systematic budget summarized in Table~\ref{tab:systematics}, we propagate each contribution as a relative uncertainty on $\sqrt{\textrm{sig}}$, which scales linearly with the inferred $\kappa$ for a fixed detected-photon sensitivity. The camera-related contribution combines the pixel gain and offset uncertainties, giving a $7.0\%$ contribution to $\sqrt{\textrm{sig}}$. We treat the optical-efficiency terms as independent multiplicative uncertainties. The power-meter calibration, ND-filter transmission, and ROI-collection terms give an optical-efficiency subtotal of $11.0\%$ in $\sqrt{\textrm{sig}}$. The ROI-collection uncertainty is dominated by the uncertainty in the $z$-direction distance between the lens and the detector. The weak-light \textit{in situ} calibration fixes the absolute camera response at $\omega_0$ and the lens transmission, so we do not assign additional independent uncertainties to these terms. The total detector-side systematic budget is thus $\leq 13.1\%$ in $\sqrt{\textrm{sig}}$.

The boost-factor uncertainty is discussed separately because it is not an uncertainty of the optical detection system itself, but of the stack response used to interpret a detected-photon sensitivity as a $\kappa$ exclusion. It is affected by the precision of TEM scanning and the non-uniformity of each dielectric layer, and is represented as the red shaded band in Fig.\,\ref{fig:boost} of the main text.

\begin{table}
\caption{\label{tab:systematics} Summary of systematic uncertainties.}
\vspace{0.2cm}
\begin{tabular}{@{}lcc@{}}
\toprule
Source                                 & Uncertainty & Contribution in $\sqrt{\textrm{sig}}$\\ \midrule
Camera response & & 7.0\% \\
~~~pixel gain & 0.2\,DN/$e^-$ &  \\
~~~pixel offset & 0.1\ \text{DN} &  \\\midrule
Optical efficiency & & 11.0\%\\
~~~powermeter & 3\% &  \\
~~~ND transmission & 2\% each &  \\
~~~ROI collection & 10\% &  \\\midrule
Total & & $\leq$13.1\%\\

\bottomrule
\end{tabular}
\end{table}

\suppsection{Illustrative multi-stack outlook}{app:outlook}
\paragraph{\underline{Four-stack reach projection.}}
The dashed red curve in Fig.~\ref{fig:limit} corresponds to the four-stack benchmark described in the main text. We use a conservative stack-response model with representative layer-thickness fluctuations of 15~nm for the thicker SiO$_2$ layers and 10~nm for the thinner TiO$_2$ layers. The reach is based on the total signal-yield sensitivity of the four modules and does not include a likelihood fit to the relative amplitudes across stack channels. The idealized comparison below is used separately to isolate the spectral-coverage gain from multiple tuned stacks.

\paragraph{\underline{Four-stack vs.  single-stack coverages.}}
To isolate the spectral-coverage gain from multiple tuned stacks at fixed total aperture, we use four ideal periodic mirror-backed 50-pair TiO$_2$/SiO$_2$ stacks, with the back side terminated by an ideal mirror.
The four design wavelengths are chosen to match common calibration lasers, 780, 632, 535, and 405~nm, corresponding to design masses 1.590, 1.962, 2.317, and 3.061~eV. The comparison is evaluated on a 1000-point grid from 1.45 to 3.10~eV. Each stack is assigned one quarter of the total aperture, and the four statistically independent responses are combined in quadrature in a toy discovery metric proportional to the stack response amplitude.

As a fixed-aperture reference, we compare this configuration with a full-area, full-exposure mirror-backed single stack centered at 632~nm. The four-stack layout exceeds this reference over 76.1\% of the sampled mass range, with a median discovery-significance gain of 2.58. Repeating the comparison with a 535-nm mirror-backed single-stack reference gives a similar result, 72.8\% of the band and a median gain of 2.73. This laser-aligned benchmark is intended as a calibration-friendly example based on ideal stack responses. Optimizing the number of modules, center wavelengths, and area allocation can further improve the coverage.

\paragraph{\underline{Cross-stack mass localization.}}
The relative-amplitude pattern across differently tuned stacks also provides a handle for localizing the mass of a candidate excess. Using the same ideal mirror-backed responses, we generate Asimov counts in the four stack channels for an injected dark-photon mass $m_0$ and fit them over the mass grid with a Poisson likelihood. A common non-negative signal normalization is profiled in the fit, so the mass information is driven mainly by the relative response pattern across the stacks rather than by the absolute signal yield.

For injections at 1.70, 1.80, 1.90, 2.00, 2.317, and 3.061~eV, the best-fit masses are 1.699, 1.800, 1.899, 2.000, 2.317, and 3.062~eV, respectively. The 1.80-eV injection gives a compact local 68\% interval from 1.798 to 1.808~eV, while the 2.00-eV injection gives two compact local 68\% regions at 2.000~eV and 2.007 to 2.012~eV. Other tested injections retain additional allowed regions, such as a secondary candidate near 2.682~eV for the 1.90-eV injection, reflecting degeneracies in the ideal periodic mirror-backed stack response. In a real candidate follow-up, these disconnected regions would be treated as alternative mass hypotheses and could guide the choice of subsequent stack tunings.
\end{document}